\documentclass[prl,twocolumn,showpacs]{revtex4}
\usepackage{amsfonts}
\usepackage{amsmath}
\usepackage{amssymb}
\usepackage{graphicx}

\DeclareMathOperator{\sh}{sh}

\DeclareMathOperator{\cth}{cth}
\DeclareMathOperator{\ctg}{ctg}
\DeclareMathOperator{\ch}{ch}
\input{tcilatex}

\begin{document}

\title{Diffusion in the general theory of relativity}
\author{Joachim Herrmann}
\affiliation{Max Born Institute, Max Born Stra\ss e 2a , D12489 Berlin, Germany}
\email{jherrman@mbi-berlin.de}
\pacs{05.10.Gg,
03.30.+p, 04.20.-q}

\begin{abstract}
The Markovian diffusion theory in the phase space is generalized within the
framework of the general theory of relativity. The introduction of moving
orthonormal frame vectors both for the position as well the velocity space
enables to bypass difficulties in the general relativistic stochastic
calculus. The general relativistic Kramers equation in the phase space is
derived both in the parametrization of phase space proper time and the
coordinate time. The transformation of the obtained diffusion equation under
hypersurface-preserving coordinate transformations is analyzed and diffusion
in the expanding universe is studied. It is shown that the validity of the
fluctuation-dissipation theorem ensures that in the quasi-steady state
regime the result of the derived diffusion equation is consistent with the
kinetic theory in thermodynamic equilibrium.
\end{abstract}

\maketitle

\section{1. Introduction}

The theory of Markovian diffusion within the framework of the relativity
theory has re-attracted considerable interest during the last decade.
Compared with the non-relativistic case, relativistic Markovian diffusion
becomes significantly more difficulty due to conceptual and technical
problems. Over the years many papers have been devoted to this issue (see
e.g. \cite{r1} - \cite{a3}), a few studied also diffusion in gravitational
fields within the theory of general relativity \cite{a1} - \cite{a3}.
However, at least up to the knowledge of the author, a general accepted
formulation of the theory of Markovian diffusion within the framework of
general relativity is still missing. Most investigations of statistical
processes in relativistic fluids in cosmology and astrophysics used
alternative physical approaches given by relativistic kinetic theory or
phenomenological relativistic thermodynamics \cite{r14}-\cite{r17}. In
particular the role of dissipative processes in the early evolution of the
Universe has been studied by using non-equilibrium thermodynamics \cite{d2}-%
\cite{d3}. The relativistic Boltzmann equation is an integral-differential
equation which is difficulty to solve in the non-equilibrium case far from
equilibrium. On the other hand, in the probabilistic approach
Fokker-Plack-type equations are differential equations which can be much
simpler solved. Besides, the probabilistic theory of diffusion processes
within the framework of general relativity exhibit a fundamental interest
and play a significant role in astrophysical and cosmological problems (see
e.g. \cite{k4} - \cite{a6} ).

A crucial factor in relativistic diffusion is the fact that the velocity
space in relativity is a hyperboloid (or a special 3-dimensional Riemannian
manifold) embedded into the 4-dimensional velocity Minkowski space. In the
derivation of the relativistic diffusion equation this specific feature has
to be taken into account in appropriate way, otherwise inconsistent results
arise. One of the main difficulties in the derivation of a consistent
Markovian diffusion equation in relativity is to handle the fundamental
Wiener process in a stochastic differential equation on non-Euclidian
manifolds in a rigorous way. There exist a well developed and rigorous
mathematical theory of stochastic differential equations and diffusion
processes in the base space on Riemannian manifolds with a definite metric
signature \cite{r24}, \cite{r25}. However, this stochastic calculus can not
be applied for manifolds with indefinite metric. Recently, the author
presented a physically motivated modification of this calculus to describe
diffusion within the framework of special relativity in the the phase space
of position and velocity. \cite{r26}. In the present paper this formalism
will be extended to a theory within the framework of general relativity in
external gravitational fields. The main aim of the paper is the derivation
of a generalized Kramers equation in external gravitational fields within
the framework of general relativity theory both in the parametrization of
the phase-space proper time and of the observer time within the approach of
spacetime decomposition. The transformation property with respect of
foliation-preserving coordinate transformations is analyzed and diffusion in
the evolving universe is studied. It is shown that the quasi-steady solution
is in agreement with the kinetic theory in thermodynamic equilibrium.

The paper is organized as follows. In Sec. 2 the principal results
pertaining to the special relativistic diffusion process is briefly
described. In Sec. 3 the formalism of Markovian diffusion in gravitational
fields within the framework of general relativity is presented both in the
parametrization of the phase-space proper time and the observer time. In
Sec. 4 the transformation property of the derived diffusion equation is
studied. As an example for the application of the formalism diffusion
processes in the expanding universe are discussed in Sec. 5. and the
conclusions are presented in Sec. 6.

\section{2. The special relativistic diffusion process}

Let us first review the principal results pertaining to the special
relativistic case \cite{r26} used as the basic for the generalization within
the frame of general relativity.

The important point in relativistic diffusion is the observation that the
velocity space in special relativity is a noncompact hyperbolic
3-dimensional Riemannian manifold embedded into the 4-dimensional velocity
Minkowski space. Using normalized velocity variables $u^{\mu }$ ($\mu
=0,1,2,3)$ the hyperbolic metric structure for a relativistic system of
massive particles is described by the relation

\begin{equation}
(u^{0})^{2}-(u^{1})^{2}-(u^{2})^{2}-(u^{3})^{2}=1.  \label{Eq.(1)}
\end{equation}%
Therefore relativistic Markovian diffusion processes can be described in a
rigorous way by using the mathematical stochastic calculus on Riemannian
manifolds, but adopted to the velocity space.

Stochastic differential equations in diffusion theory with continuous
pathway are defined by the fundamental Wiener process W$^{a}(t).$ On a
Riemannian manifold the fundamental Wiener process is difficult to handle.
The key idea in the mathematical concept of diffusion on general Riemannian
manifolds with definite metric signature is to define a stochastic process
on the curved manifold using the fundamental Wiener process each component
of which is a process in the Euclidian space R$^{d}$ \cite{r24},\cite{r25}.
The tangent space of a Riemannian manifold is endowed with an Euclidian
structure and therefore we can move the manifold in the tangent space by
construction of a parallel translation along the stochastic curve with the
help of the orthonormal frame vectors $e_{a}^{{}}=$\ $e_{a}^{i}(\mathbf{x}%
)\partial _{i}$ (i,a=1...d) and the Christoffel connection coefficients $%
\Gamma _{ib}^{j}$, $\mathbf{x}=(x_{1}...x_{d}),\partial _{i}=\partial
/\partial x^{i}.$ In local coordinates on a Riemannian manifold the
infinitesimal motion of a smooth curve c$^{i}(t)$ in M$^{d}$ is that of $%
\gamma ^{i}(t)$ in the tangent space by using a parallel transformation: $%
dc^{i}=e_{a}^{i}(\mathbf{x})d\gamma ^{a}$ and $de_{a}^{i}(\mathbf{x}%
)=-\Gamma _{ml}^{i}e_{a}^{l}dc^{m}$. Therefore a \ random curve in the
stochastic mathematical calculus on Riemannian manifolds in the position
space can be defined in the same way by using the canonical realization of a
d-dimensional Wiener process (defined in the Euclidian space) and
substituting $d\gamma ^{a}\rightarrow dW^{a}(t)$. Thus the stochastic
differential equations describing diffusion on a Riemannian manifold in the
orthonormal frame bundle $O(M)$ with coordinates $O(M)$ =$%
\{x^{i},e_{a}^{i}\} $ are given by \cite{r24},\cite{r25}

\begin{eqnarray}
dx^{i}(\tau ) &=&e_{a}^{i}(\mathbf{\tau })\circ dW_{\tau }^{a}+b^{i}(\tau
)d\tau ,  \label{equ.1} \\
de_{a}^{i}(\mathbf{\tau }) &=&-\Gamma _{ml}^{i}e_{a}^{l}\circ dx^{m}(\tau ).
\notag
\end{eqnarray}%
Here $\delta ^{ab}e_{a}^{i}(\mathbf{x}(\tau ))e_{b}^{j}(\mathbf{x}(\tau
))=g^{ij},\partial _{i}e_{a}^{j}=-\Gamma _{ik}^{j}e_{a}^{k}$, $g^{ij}$ is
the Riemannian metric and $\delta ^{ab}$ the flat Euclidian metric where $%
\delta ^{ab}$ is the Kronecker symbol.

\bigskip The general mathematical model given by Equ.(2) describes diffusion
in the position space. It can not be generalized to pseudo-Riemannian
manifolds in the general relativity theory. But Equ.(2) exhibit a
significant difference to the non-relativistic Langevin equation in physics,
where the stochastic force acts directly only on the change of the velocity
and not on the position coordinates. Therefore the mathematical stochastic
calculus for Riemannian manifolds can be applied with an appropriate
modification adopting it to the velocity space. This requires the
introduction of a moving velocity frame $E_{a}^{i}(\tau ).$ A relativistic
generalization of the Langevin equations can be defined in the fiber bundle
space $F(M_{L})=\{x^{i},u^{i},$ $E_{a}^{i}\}$ by \cite{r26}
\begin{eqnarray}
dx^{i}(\tau ) &=&u^{i}(\tau )d\tau ,  \label{Eq.2} \\
du^{i}(\tau ) &=&E_{a}^{i}(\tau )\circ dW^{a}(\tau )+F^{i}(\tau )d\tau ,
\notag \\
dE_{a}^{i}(\tau ) &=&-\gamma _{ml}^{i}(\mathbf{u})E_{a}^{l}(\tau )\circ
du^{m}(\tau ).  \notag
\end{eqnarray}%
\ Here $\tau $ is an evolution parameter along the world lines of the
particles which can be chosen as the proper time. The laboratory time $%
t=\tau u^{0}$/$c$ is a function of the proper time $\tau $ and $u^{0}$ which
here and below is defined by $u^{0}=[1+(u^{1})^{2}+(u^{2})^{2}+(u^{3})^{2}]^{%
\frac{1}{2}}$. $\gamma _{ml}^{i}(\mathbf{u})$ are the Christoffel connection
coefficients on the hyperboloid and $F^{i}$=$K^{i}/m$, where $K^{i}$ are the
spatial components of the 4-force ($i=1,2,3),$ m is the rest mass of the
particles and the $a,b$ numbers the moving frame vectors $E_{a}$ in the
hyperbolic velocity space ($a,b=1,2,3$). The stochastic force is described
by the fundamental Wiener process with $\langle W^{a}\rangle =0$ and the
correlator \ $\langle W^{a}(\tau )W^{b}(\tau +s)\rangle =Ds\delta _{ab}$, is
defined by an empirical diffusion constant $D$ which is independent on the
velocity. Stochastic integrals related with Equ.(3) are defined in the
Stratonovich calculus denoted by the symbol $\circ .$ Since the manifold on
the hyperboloid is embedded into the Minkowski space, metric $G_{ij}(u)$ and
connection coefficients $\gamma _{jk}^{i}(\mathbf{u})$ in the velocity space
are given by
\begin{eqnarray}
G_{ij}(\mathbf{u}) &=&\delta _{ij}-(u^{i}u^{j})/(u^{0})^{2},
\label{Equ.(3a)} \\
\gamma _{jk}^{i}(\mathbf{u}) &=&-u^{i}G_{jk}.  \notag
\end{eqnarray}%
Associated with the diffusion process described by Equ.(3) is a diffusion
generator $\mathbf{A}_{F(M_{L})}$

\begin{equation}
\mathbf{A}_{F(M_{L})}=\frac{D}{2}\delta ^{ab}H_{a}H_{b}+H_{0},
\label{Eq.(3)}
\end{equation}%
where the fundamental horizontal vector fields $H_{a}$ and $H_{0}$ on the
fiber bundle $\emph{F}(M_{L})$ are given by
\begin{eqnarray}
H_{a} &=&E_{a}^{i}\frac{\partial }{\partial u^{i}}-\gamma _{ml}^{i}(\mathbf{u%
})E_{a}^{l}\ E_{b}^{m}\frac{\partial }{\partial E_{b}^{i}},  \label{Eq.(4)}
\\
H_{0} &=&u^{i}\frac{\partial }{\partial x^{i}}+F^{i}\frac{\partial }{%
\partial u^{i}}-\gamma _{ml}^{i}(\mathbf{u})E_{a}^{l}(\tau )F^{m}\frac{%
\partial }{\partial E_{a}^{i}}.  \notag
\end{eqnarray}%
We project the stochastic curve from the fiber space\ $F(M_{L})$ with
coordinates $\mathbf{r}=$\{$x^{i},u^{i},E_{a}^{i}\}$ to the phase space with
coordinates \{$x^{i}$,$u^{i}\}:\mathbf{A}_{F(M_{L})}f(\mathbf{r})=\mathbf{A}%
_{P}f(\mathbf{x},\mathbf{u,0})$, where the diffusion generator in the phase
space $\mathbf{A}_{P}$ is given by
\begin{equation}
\mathbf{A}_{P}=\frac{D}{2}\delta ^{ab}E_{a}^{i}\frac{\partial }{\partial
u^{i}}E_{b}^{j}\frac{\partial }{\partial u^{j}}+u^{i}\partial /\partial
x^{i}+F^{i}\partial /\partial u^{i}.  \label{Eq.(5)}
\end{equation}

The generator $\mathbf{A}_{P}$ describes how the expected value $\varphi
(\tau ,\mathbf{x})=E^{x}[f(\mathbf{X}_{\tau })]$ of any smooth function $f$
evolve in time and satisfies the following equation:

\begin{equation}
\frac{\partial }{\partial \tau }\varphi (\tau ,\mathbf{x,u})=\mathbf{A}%
\varphi (\tau ,\mathbf{x,u}),  \label{Eq.(6)}
\end{equation}%
with $\varphi $(0,$\mathbf{u}$)=$f$($\mathbf{x}$). Equ. (8) is a Kolmogorov
backward equation. A Fokker-Planck equation (or forward Kolmogorov equation)
describes how the probability density function $\phi (\tau ,\mathbf{x})$
evolves with time and is determined by the adjoint of the diffusion operator
$\mathbf{A}_{P}^{\ast }.$ Therefore the special relativistic diffusion
equation in the phase space takes the form \cite{r26}:

\begin{equation}
\frac{\partial \phi }{\partial \tau }=-u^{i}\frac{\partial \phi }{\partial
x^{i}}-div_{u}(\mathbf{F}\phi )+\frac{D}{2}\Delta _{u}\phi ,  \label{Eq.(7)}
\end{equation}%
where $\Delta _{u}$ is the Laplace Beltrami Operator of the hyperbolic
velocity space given by

\begin{eqnarray}
\Delta _{u} &=&G^{ij}\frac{\partial ^{2}}{\partial u^{i}\partial u^{j}}%
-G^{ij}\gamma _{ij}^{k}\frac{\partial }{\partial u^{k}}  \label{Eq.(8)} \\
&=&\frac{1}{\sqrt{G}}\frac{\partial }{\partial u^{i}}(\sqrt{G}G^{ij}\frac{%
\partial }{\partial u^{j}}),  \notag
\end{eqnarray}%
and the corresponding divergence operator is given by

\begin{equation}
div_{u}(\mathbf{F}\phi )=\frac{1}{\sqrt{G}}\frac{\partial }{\partial u^{i}}(%
\sqrt{G}F^{i}\phi ),  \label{Eq.(9)}
\end{equation}%
with $G=\det \{G_{ij}\}=(u_{0})^{-2}$ and $G^{ij}=\delta ^{ij}+u^{i}u^{j}.$

\section{3. General relativistic diffusion equation}

The special relativistic formalism briefly described in section 2 can be
extended to a theory within the framework of general relativity. Besides
orthonormal frame vectors in the velocity space in the generalization of his
formalism the introduction of orthonormal frame vectors in the position
space is required.

In the case of general relativity the four-velocity $v^{\mu }$ of massive
particles in the presence of a gravitational field with a metric $g_{\mu \nu
}(x)$ satisfies the condition

\begin{equation}
g_{\mu \nu }(x)v^{\mu }v^{\nu }=1,  \label{Eq. (10a)}
\end{equation}%
depending not only on the velocities $v^{\nu }$ but also on the coordinates x%
$^{\mu }.$ The complication arising by this fact can be bypassed by using
the orthonormal frame vectors $e_{M}=$ $e_{M}^{\mu }(x)\partial _{\mu }$ in
the position space, where the subscript $M,N$ numbers the vectors ($%
M,N=0,1,2,3$), $\mu $ their components in the coordinate basis $e_{\mu }=$ $%
\partial _{\mu }$ ($\mu =0,1,2,3)$ and $x=\{x^{0},x^{1},x^{2},x^{3}).$ The
frame vectors satisfy the condition

\begin{eqnarray}
\eta ^{MN}e_{M}^{\mu }(x)e_{N}^{\nu }(x) &=&g^{\mu \nu },  \label{Eq. (11a)}
\\
g_{\mu \nu }e_{M}^{\mu }(x)e_{N}^{\nu }(x) &=&\eta _{MN},  \notag
\end{eqnarray}%
where $\eta _{MN}=diag(-1,1,1,1)$ is the metric of the Minkowski \ space.
The dual basis of the frame fields $e_{M}$ are cotangent frame 1-forms $\
\theta ^{M}=\theta _{\mu }^{M}dx^{\mu }$ satisfying the orthogonality
relation
\begin{equation}
e_{M}^{\mu }(x)\theta _{\nu }^{M}(x)=\delta _{\nu }^{\mu }.
\label{Eq. (12a)}
\end{equation}%
Each vector refer to the coordinate basis $e_{\mu }=$ $\partial _{\mu }$ is
assigned a vector refer to the frame basis $e_{M}$ according to the rule
\begin{equation}
v^{\nu }=e_{M}^{\nu }(x)v^{M}.  \label{Equ. (14b)}
\end{equation}%
Correspondingly, we find for $v^{M}$

\begin{equation}
v^{M}=\theta _{\nu }^{M}(x)v^{\nu }.  \label{Eq. (14a)}
\end{equation}%
By using the orthogonal moving frames of the pseudo-Riemannian manifold in
the position space the relation (12) gets the same form as in the
special-relativistic case in Equ. (1):
\begin{eqnarray}
g_{\mu \nu }(x)v^{\mu }v^{\nu } &=&g_{\mu \nu }(x)e_{M}^{\mu }(x)e_{N}^{\nu
}(x)v^{M}v^{N}  \label{Eq. (15a)} \\
&=&\eta _{MN}v^{M}v^{N}=1.  \notag
\end{eqnarray}%
The covariant derivative in the natural frame is defined as usual,

\begin{equation}
\nabla _{\mu }v^{\nu }=\partial _{\mu }v^{\nu }+\Gamma _{\mu \rho }^{\nu
}v^{\rho },  \label{Eq. (16a)}
\end{equation}%
where $\Gamma _{\mu \rho }^{\nu }$ are the Christoffel connection
coefficients (or coordinate connections). The covariant derivative of
vectors refer to the frame basis can be written as:%
\begin{equation}
\nabla _{\mu }v^{M}=\partial _{\mu }v^{M}+\Omega _{\mu N}^{M}v^{N},
\label{Eq. (17a)}
\end{equation}%
where $\Omega _{\mu N}^{M}$ are called coefficients of spin connection (or
frame connection). The relation between these two kind of connections arise
from the metric compatibility condition, which here can be expressed by $%
\nabla _{\mu }\theta _{\nu }^{M}=0,\nabla _{\mu }e_{M}^{\nu }=0.$ From this
condition one gets \cite{r27}, \cite{r28}%
\begin{equation}
\Omega _{\mu N}^{M}(x)=\theta _{\nu }^{M}(x)\Gamma _{\mu \rho }^{\nu
}e_{N}^{\rho }(x)+\theta _{\nu }^{M}(x)\partial _{\mu }e_{N}^{\nu }(x).
\label{Eq. (18a)}
\end{equation}

The parallel transport of the components of the 4-vector $v^{M}$ in the
moving frame $e_{A}$ and the line elements\ $dx^{\mu }$ are given by
\begin{eqnarray}
dv^{M} &=&-\Omega _{\mu N}^{M}(x)v^{N}dx^{\mu },  \label{Eq.(20)} \\
dx^{\mu } &=&e_{M}^{\mu }(x)v^{M}d\tau .  \notag
\end{eqnarray}

\bigskip As seen in Equ.(21)\ the change of the spatial components of the
velocity $v^{m}$ of a particle is determined by the gravitational force
described by the spin connection coefficients $\Omega _{\mu N}^{m}(x).$ In
addition one has to take into account that the velocity is changed by a
stochastic force $F_{noise}^{a}$ driven by the Wiener process $dW^{a}(t).$
In order to avoid difficulties in the description of the Wiener process on
the velocity hyperbolic manifold the mathematical stochastic calculus on
Riemannian manifolds has to be applied where the Wiener process is moved
along the orthonormal frames $E_{a}^{i}(u)$ \cite{r26} \ in the
three-dimensional hyperbolic velocity space defined by the relation

\begin{equation}
\sum\limits_{a=1}^{3}E_{a}^{m}E_{a}^{n}=G^{mn},  \label{Eq. (21a)}
\end{equation}%
or equivalently%
\begin{equation}
G_{mn}E_{a}^{m}E_{b}^{n}=\delta _{ab},  \label{Eq. (22a)}
\end{equation}%
where $G_{ij}$ is the Riemannian metric of the hyperbolic velocity space and
$G^{ij}$ the inverse matrix of $G_{ij}$. The metric and the Christoffel
coefficients in the velocity space are defined in Equ. (4). The
infinitesimal motion of the velocity $v^{m}(\tau )$ in phase space can be
described by that of $u^{a}(\tau )$ ($a=1,2,3)$ in the moving velocity frame
$E_{a}^{m}$ by using the parallel transformation
\begin{eqnarray}
dv^{m} &=&E_{a}^{m}du^{a}  \label{Eq. (23a)} \\
dE_{a}^{m}(\tau ) &=&-\gamma _{nl}^{m}(\mathbf{v})E_{a}^{l}(\tau )dv^{n}
\notag
\end{eqnarray}%
A random curve in the phase-space can be defined as in section 2 by using
the Wiener process substituting $du^{a}\rightarrow dW^{a}(\tau ).$ Therefore
analogous as in Equ.(3) in a consistent description of Markovian diffusion
in general relativity the noise force is described by $%
F_{noise}^{m}=E_{a}^{m}(\tau )\circ dW^{a}.$ A remarkable feature of
Markovian diffusion on a Riemannian manifold is the supposition that for the
diffusion coefficients only the orthonormal frame coefficients $E_{b}^{a}$
are admissible which are directly related to the hyperbolic geometry of the
velocity space. In contrast, on Euclidian manifolds a much more general
class of diffusion coefficients are permitted. Therefore generalizing
Equ.(3) the stochastic differential equations describing diffusion in a
gravitational and external force fields (general relativistic Langevin
equations) are given by:

\begin{eqnarray}
dx^{i} &=&e_{M}^{i}(x)v^{M}d\tau ,  \label{Eq.(25)} \\
dv^{m} &=&E_{a}^{m}(\tau )\circ dW^{a}-\Omega _{\mu N}^{m}(x)e_{M}^{\mu
}(x)v^{N}v^{M}d\tau +  \notag \\
&&+F_{ex}^{m}d\tau ,  \notag \\
dE_{a}^{m}(\tau ) &=&-\gamma _{nl}^{m}(\mathbf{v})E_{a}^{l}(\tau )\circ
dv^{n}.  \notag
\end{eqnarray}%
Here an additional external force $F_{ex}^{\mu }$=$K_{ex}^{\mu }/m$ is taken
into account where $K_{ex}^{\mu }$ are the components of the external
4-force in the coordinate frame and m is the rest mass of the particles. In
the moving frame this force is expressed by $F_{ex}^{m}=\theta _{\mu
}^{m}F_{ex}^{\mu }$. The Christoffel connection coefficients on the
hyperboloid $\gamma _{nl}^{m}(\mathbf{v})$ are defined in Equ. (4) and $\tau
$ is a parameter defined along the world line of the particles, which can be
chosen as the phase-space proper time.

Sufficient and necessary conditions for the existence and uniqueness of the
stochastic differential equation (25) are that the drift and diffusion
coefficients satisfy the uniform Lipschitz condition and the stochastic
process $\mathbf{X}$($\tau $)=\{$\mathbf{x}$($\tau $), $\mathbf{v}$($\tau $%
)\} is adapted to the Wiener process W$^{a}$($\tau )$ , that is, the output
X($\tau _{2}$) is a function of W$^{a}$($\tau _{1})$ up to that time ($\tau
_{1}\leq \tau _{2})$.

\bigskip The diffusion operator $\mathbf{A}_{F(M)}$ in the fiber bundle $%
F(M_{L})$ with coordinates $\mathbf{r}=$\{$x^{i},v^{m},E_{a}^{m}\}$ for the
stochastic process described by Equ. (25) is defined as in Equ.(5) with
horizontal vector fields $H_{a}$ and $H_{0}$ derived analogous as in Sec.2:
\begin{eqnarray}
H_{a} &=&E_{a}^{m}\frac{\partial }{\partial v^{m}}-\gamma _{nl}^{m}(\mathbf{v%
})E_{a}^{l}\ E_{b}^{n}\frac{\partial }{\partial E_{b}^{m}},  \label{Eq.(26)}
\\
H_{0} &=&e_{M}^{i}(x)v^{M}\frac{\partial }{\partial x^{i}}-\Omega _{\mu
N}^{m}(x)e_{M}^{\mu }(x)v^{N}v^{M}\frac{\partial }{\partial v^{m}}+  \notag
\\
&&+F_{ex}^{m}\frac{\partial }{\partial v^{m}}-\gamma _{nl}^{m}(\mathbf{v}%
)E_{a}^{l}\ F^{n}\frac{\partial }{\partial E_{a}^{m}}.  \notag
\end{eqnarray}%
The operator $\mathbf{A}_{F(M)}$ can be projected to the phase space with
coordinates $\mathbf{r}=$\{$x^{i}$,$v^{m}\}$ by $\mathbf{A}_{F(M)}f(\mathbf{r%
})=\mathbf{A}_{P}f(\mathbf{x},\mathbf{u})$, where the diffusion generator in
the phase space $\mathbf{A}_{P}$ is given by
\begin{eqnarray}
\mathbf{A}_{P} &=&\frac{D}{2}\delta ^{ab}E_{a}^{m}\frac{\partial }{\partial
v^{m}}E_{b}^{n}\frac{\partial }{\partial v^{n}}+e_{M}^{i}(x)v^{M}\frac{%
\partial }{\partial x^{i}}+  \notag \\
&&+F^{m}\frac{\partial }{\partial v^{m}}\mathbf{.}  \label{Equ. /27b)}
\end{eqnarray}%
Here the first term contains the Laplace-Beltrami operator $\Delta
_{v}=\delta ^{ab}E_{a}^{m}\frac{\partial }{\partial v^{m}}E_{b}^{n}\frac{%
\partial }{\partial v^{n}}$ in the hyperbolic velocity space. The force $%
F^{m}$ is composed of the gravitational force and the external force $%
F_{ex}^{m}$: $F^{m}=$ $-\Omega _{\mu N}^{m}(x)e_{M}^{\mu
}(x)v^{N}v^{M}+F_{ex}^{m}.$ The backward Kolmogorov equation for the
stochastic process described by Equ.(25) is defined by

\begin{equation}
\frac{\partial }{\partial \tau }\varphi (\tau ,\mathbf{x,v})=\mathbf{A}%
_{P}\varphi (\tau ,\mathbf{x,v}).  \label{Eq.(29)}
\end{equation}%
The coresponding Fokker-Planck equation in phase space (general
relativistic Kramers equation) within the frame of general
relativity is defined by the adjoint of the operator
$\mathbf{A}_{P}.$ Since the Laplace-Beltrami operator is
self-adjoint $\Delta _{v}=\Delta _{v}^{+}$ it takes the form:

\begin{equation}
\frac{\partial \Phi }{\partial \tau }=-v^{M}div_{x}(\mathbf{e}_{M}(x)\Phi )\
-div_{v}(\mathbf{F}\Phi )+\frac{D}{2}\Delta _{v}\Phi ,  \label{Eq.(30)}
\end{equation}%
where $\Delta _{v}$ is the Laplace-Beltrami Operator in the
hyperbolic velocity space

\begin{eqnarray}
\Delta _{v} &=&G^{mn}\frac{\partial ^{2}}{\partial v^{m}\partial v^{n}}%
+G^{mn}\gamma _{mn}^{l}\frac{\partial }{\partial v^{l}}  \label{Eq.(31)} \\
&=&\frac{1}{\sqrt{G}}\frac{\partial }{\partial v^{m}}(\sqrt{G}G^{mn}\frac{%
\partial }{\partial v^{n}}).  \notag
\end{eqnarray}%
The divergence operator in the position space is given by
\begin{equation}
div_{x}(\mathbf{e}_{M}(x)\Phi )=\frac{1}{\sqrt{g}}\frac{\partial }{\partial
x^{i}}(\sqrt{g}e_{M}^{i}(x)\Phi ),  \label{Eq.(32)}
\end{equation}%
and in the velocity space by

\begin{equation}
div_{v}(\mathbf{F}\Phi )=\frac{1}{\sqrt{G}}\frac{\partial }{\partial v^{m}}(%
\sqrt{G}F^{m}\Phi ),  \label{Eq.(33)}
\end{equation}%
with $G=\det \{G_{ij}\},g=\det \{g_{ij}\}$.

Equ.(29) is the diffusion equation in the parametrization of the phase-space
proper time within the frame of general relativity for the probability
density function $\Phi =\phi (\tau ;\mathbf{x},\mathbf{v})$ or the the
transition probability $\Phi (\mathbf{x},\mathbf{v,\tau }\mid \mathbf{x}_{0},%
\mathbf{v}_{0},0)$.

For the solution of the relativistic diffusion equation it is convenient to
introduce the hyperbolic coordinate system for the 4-velocity defined by $%
v^{1}=\sh\alpha \sin \vartheta \cos \varphi ,$ $v^{2}=\sh\alpha \sin
\vartheta \sin \varphi ,$ $v^{3}=\sh\alpha \cos \vartheta $ and $v^{0}=\ch%
\alpha $. We denote the velocities in the non-Cartesian coordinates by $%
\overline{v}^{1}=\alpha $,$\overline{v}^{2}=\theta $,$\overline{v}%
^{3}=\varphi $ $,a=1,2,3.$ The metric in this coordinates are simply to
calculate and are given by $G_{11}=1,$ $G_{22}=sh^{2}\alpha
,G_{33}=sh^{2}\alpha \sin ^{2}\vartheta \ $ and $G_{ij}=0$ for $i\neq
j,G=sh^{4}\alpha \sin ^{2}\vartheta .$ With the given metric the Laplace
Beltrami Operator $\Delta _{u}$ takes the form

\begin{eqnarray}
\Delta &=&\frac{\partial ^{2}}{\partial \alpha ^{2}}+2\cth\alpha \frac{%
\partial }{\partial \alpha }-  \label{Eq.(34)} \\
&&-\frac{1}{(\sh\alpha )^{2}}\left( \frac{\partial ^{2}}{\partial \vartheta
^{2}}+\ctg\vartheta \frac{\partial }{\partial \vartheta }+\frac{1}{(\sin
\vartheta )^{2}}\frac{\partial ^{2}}{\partial \varphi ^{2}}\right)  \notag
\end{eqnarray}%
%
%
%
%
%
%
%
%
%
%
%
%
%
%
%
%
%
%
%
%
%
%
%
%
%
%
%
%
%
%
%
%
%
%
%
%
%
%
%
%
%
%
%
%
%
%
%
%
%
%
%
%
%
%
%
%
%
%
%
%
%
%
%
%
%
%
%
%
%
%
%
%
%
%
%
%
%
%
%
%
%
%
%
%
%
%
%
%
%
%
%
%
%
%
%
%
%
%
%
%
%
%
%
%
%
%
%
%
%
%
%
%
%
%
%
%
and%
\begin{eqnarray}
\text{div}_{v}(\mathbf{F}\Phi ) &=&(\sh\alpha )^{-2}\frac{\partial }{%
\partial \alpha }\left( (\sh\alpha )^{2}F^{\alpha }\Phi \right) -
\label{Equ. (34b)} \\
&&-(\sh\alpha )^{-1}(\sin \vartheta )^{-1}\frac{\partial }{\partial
\vartheta }\left( \sin \vartheta F^{\vartheta }\Phi \right) -  \notag \\
&&-(\sh\alpha )^{-1}(\sin \vartheta )^{-1}\frac{\partial }{\partial \varphi }%
(F^{\varphi }\Phi ),  \notag
\end{eqnarray}%
is the divergence operator. Here the force components in the hyperbolic
coordinate system $F^{\alpha },F^{\vartheta },F^{\varphi }$ are related with
$F^{i}$ by $F^{\alpha }=(\ch\alpha )^{-1}\ [\sin \vartheta (\cos \varphi
F^{1}+\sin \varphi F^{2})+\cos \vartheta F^{3}]$, $F^{\vartheta }=(\sh\alpha
)^{-1}\ [\cos \vartheta (\cos \varphi F^{1}+\sin \varphi F^{2})-\sin
\vartheta F^{3}]\ $and $F^{\varphi }=(\sh\alpha )^{-1}\ (\sin \vartheta
)^{-1}[-\sin \varphi F^{1}+\cos \varphi F^{2}].$

The relativistic diffusion equation (29) is parameterized in terms of the
phase-space proper time $\tau .$ But it is more convenient to parameterize
the stochastic process alternatively in terms of the coordinate time because
the gravitational fields and the external force fields are given in this
parametrization. In general relativity the observer time with the
infinitesimal element $dx^{0}=e_{M}^{0}(x)v^{M}d\tau $ is a function of the
proper time $\tau $ and the space and velocity variables. This general
definition introduces difficulties in the diffusion formalism. In order to
avoid such problem and to simplify the derivation we describe diffusion in a
frame system in which the time-like components of the frames $e_{m}(x)$
vanish: $e_{m}^{o}(x)=0.$ This condition can be achieved in general if we
impose three frame subsidiary conditions and remove a part of the six frame
arbitraries. In particular, the condition $e_{m}^{o}(x)=0$ is introduced in
the 1+3 spacetime slicing in the Arnovitt-Deser-Misner (ADM) formalism \cite%
{r29} of the hamiltonian formulation of gravity. In the ADM treatment the
space-time manifold is split into a one-parameter family of space-like
hypersurfaces $\Sigma (t)$ parameterized by a time-like function $t$ or as a
foliation of the hypersurfaces $t=const.$

With vanishing frame components $e_{m}^{o}(x)=0$ the proper time is given by
$dx^{0}=(g_{00})^{\frac{1}{2}}v^{0}d\tau $ where $v^{0}$ is defined by the
relation (17). The diffusion equation in the parametrization of the observer
time can be derived from the stochastic differential equation (25) \ using
the mathematical theorem of random time change in stochastic differential
equations (see e.g. \cite{r24}, \cite{r25}). The proper time $\tau $ is
related with $x^{0}$ by the random transformation
\begin{equation}
\tau =\int\limits_{0}^{x^{0}}N(s)(v^{0}(s))^{-1}ds,  \label{Eq. (36a)}
\end{equation}%
with $N=(g_{00})^{-\frac{1}{2}}.$ Since $\tau $ depends only on the
stochastic events $v^{i}$ earlier than $x^{0}$ this random time change is an
adapted process and therefore the time change of an Ito integral is again an
Ito integral, but driven by a different Wiener process $d\widetilde{W}%
(x^{0})=dW(\tau )N^{-\frac{1}{2}}(v^{0})^{1/2}$ \cite{r24}, \cite{r25}. This
rule for a random time change is valid within the Ito calculus. Using the
transformation rules of an Ito integral into a Stratonovich integral and $%
d\tau =(v^{0})^{-1}Ndx^{0}$ the relativistic Langevin equation (25) can be
rewritten in the parametrization with $x^{0}$ as follows%
\begin{eqnarray}
dx^{i}(x^{0}) &=&e_{M}^{i}(x)v^{M}(v^{0})^{-1}Ndx^{0},  \notag \\
dv^{m}(x^{0}) &=&E_{a}^{m}(v^{0})^{-1/2}N^{\frac{1}{2}}\circ d\widetilde{%
W^{a}}(x^{0})-  \label{Equ. (36b)} \\
&&-\frac{D}{2}\delta ^{ab}E_{a}^{m}E_{b}^{n}N(v^{0})^{-1/2}\cdot   \notag \\
&&\frac{\partial }{\partial v^{n}}%
((v^{0})^{-1/2})dx^{0}+F^{m}(v^{0})^{-1}Ndx^{0},  \notag \\
dE_{a}^{m}(x^{0}) &=&-\gamma _{nl}^{m}(\mathbf{v})E_{a}^{l}\circ
dv^{n}(x^{0})  \notag
\end{eqnarray}%
Note that in Equ. (36 ) an additional drift term proportional to the
diffusion constant $D$ arise which originates from the transformation rule
from the Ito to the Stratonovich calculus for random time changes. The
diffusion operator $\mathbf{A}_{F(M)}$ in the fiber bundle $F(M_{L})$ with
coordinates $\mathbf{r}=$\{$x^{i},v^{m},E_{a}^{m}\}$ for the stochastic
process described by Equ. (25) is defined as in Equ.(5) with horizontal
vector fields $H_{a}$ and $H_{0}$ now given by
\begin{eqnarray}
H_{a} &=&E_{a}^{m}(v^{0})^{-1/2}N^{\frac{1}{2}}\frac{\partial }{\partial
v^{m}}-  \label{Equ.(37c)} \\
&&-\gamma _{nl}^{m}(\mathbf{v})E_{a}^{l}E_{b}^{n}(v^{0})^{-1/2}N^{\frac{1}{2}%
}\frac{\partial }{\partial E_{b}^{m}}  \notag \\
H_{0} &=&\{-\frac{D}{2}\delta ^{ab}E_{a}^{m}E_{b}^{n}N(v^{0})^{-1/2}\frac{%
\partial }{\partial v^{n}}((v^{0})^{-1/2})+  \notag \\
&&+(v^{0})^{-1}NF^{m}\}\frac{\partial }{\partial v^{m}}%
+(v^{0})^{-1}Ne_{M}^{i}(x)v^{M}\frac{\partial }{\partial x^{i}}-  \notag \\
&&-(v^{0})^{-1}N\gamma _{nl}^{m}(\mathbf{v})E_{a}^{l}\{F^{n}+  \notag \\
&&+\frac{D}{2}\delta ^{bc}E_{b}^{n}E_{c}^{k}\frac{\partial }{\partial v^{k}}%
((v^{0})^{-1/2})\}\frac{\partial }{\partial E_{a}^{m}}  \notag \\
&&  \notag
\end{eqnarray}%
with the gravitational and the external force $F^{m}=$ $-\Omega _{\mu
N}^{m}(x)e_{M}^{\mu }(x)v^{N}v^{M}+F_{ex}^{m}.$ The operator $\mathbf{A}%
_{F(M)}$ can be projected to the phase space with coordinates $\mathbf{r}=$\{%
$x^{i}$,$v^{i}\}$ by $\mathbf{A}_{F(M)}f(\mathbf{r})=\mathbf{A}_{P}f(\mathbf{%
x},\mathbf{v})$, where the diffusion generator in the phase space $\mathbf{A}%
_{P}$ is given by
\begin{eqnarray}
\mathbf{A}_{P} &=&\frac{D}{2}(v^{0})^{-1}N\delta ^{ab}E_{a}^{i}\frac{%
\partial }{\partial v^{i}}E_{b}^{j}\frac{\partial }{\partial v^{j}}
\label{Eq. (39a)} \\
&&+e_{M}^{i}(x)v^{M}(v^{0})^{-1}N\frac{\partial }{\partial x^{i}}%
+F^{m}(v^{0})^{-1}N\frac{\partial }{\partial v^{m}}\mathbf{\ }  \notag
\end{eqnarray}

The general relativistic Kramers equation in the parametrization of the
observer time $x_{0}$ can be derived analogous as above and is written as
follows
\begin{equation}
N^{-1}v^{0}\frac{\partial \Phi }{\partial x^{0}}=-v^{M}div_{x}(\mathbf{e}%
_{M}(x)\Phi )-div_{v}(\mathbf{F}\Phi )+\frac{D}{2}\Delta _{v}\Phi .
\label{11a}
\end{equation}

\bigskip As seen in the limit of special relativity the left side and the
first term o.r.s. of Equ.(39) can be identified with the covariant
expression $v^{\mu }\partial /\partial x^{\mu },$while the other terms are
identical with corresponding terms in Equ.(29).

The equation (39) is the main result of the present paper and represents the
generalization of the Kramers equation within the frame of general
relativity in the parametrization of the observer time for the probability
density function $\Phi =\phi (x_{0};\mathbf{x},\mathbf{v})$ with the initial
condition $\phi (x_{0}=0;\mathbf{x},\mathbf{v})=\phi _{0}(\mathbf{x},\mathbf{%
v})$. The transition probability is determined by the same equation but is
defined by the initial condition $\Phi (\mathbf{x},\mathbf{v,0}\mid \mathbf{x%
}_{0},\mathbf{v}_{0},0)=(Gg)^{-1/2}\delta (v^{1}-v_{0}^{1}$)\ \ $\delta
(v^{2}-v_{0}^{2})\delta (v^{3}-v_{0}^{3}$)$\delta (x_{1}-x_{1}^{0}$)$\delta
(x_{2}-x_{2}^{0}$)$\delta (x_{3}-x_{3}^{0}$). For an external
electromagnetic field $F_{\mu \nu }$ the normalized force $F_{ex}^{m}$ in
the moving frame is $F^{m}=e\theta _{\mu }^{m}g^{\mu \rho }F_{\rho \nu
}e_{N}^{\nu }(x)v^{N}$.

\section{4. Coordinate transformations}

In the general relativistic framework, the invariance of the physical laws
with respect of general coordinate transformations is one of the most
fundamental property. In the frame of special relativity the probability
density function $\phi (\tau ;\mathbf{x},\mathbf{v})$ in the phase space is
Lorentz invariant; i.e. it fulfills the condition

\begin{equation}
\phi `(\tau `,\mathbf{x`},v\mathbf{`})=\phi (\tau ,\mathbf{x},\mathbf{v}),
\label{Eq.(39)}
\end{equation}%
where the variables $\mathbf{x`},\mathbf{v`}$ are related with $\mathbf{x},%
\mathbf{v}$ by a Lorentz transformation. Note that in contrast the particle
density and the current density (i.e. the integrals over the velocities)
transform like a four-vector. As shown in \cite{r26}\ the special
relativistic equation (9) suffices this condition and is invariant with
respect to a Lorentz transformation. In the general theory of relativity
this invariance should be valid for general coordinate transformations.
However in the derivation of Equ.(39) using the parametrization of the
observer time we have taken advantage of the freedom in the choice of the
orthonormal frame components and used the condition $e_{m}^{o}(x)=0$ or
correspondingly a hypersurface foliation. In this frame a coordinate
independent notion of time is demanded, and therefore we have to consider a
foliation-preserving diffeomorphism described by the general space
coordinate transformations

\begin{equation}
x`^{i}=f^{i}(x^{j},t),  \label{Eq. (43a)}
\end{equation}%
which preserve the hypersurface geometry. The hypersurface foliation is also
preserved under an arbitrary time rescaling%
\begin{equation}
t^{\prime }=f^{0}(t)  \label{Equ. (44a)}
\end{equation}%
which do not affect the hypersurfaces. General covariance will then become a
hidden feature similar as in the hamiltonian formulation of gravity in the
ADM formalism, but the underlying invariance of general relativity is intact
and general coordinate transformation still map solutions into solutions.

\bigskip Let us study the transformation property of the diffusion equation
(39) in the parametrization of the observer time. Since the moving 1-forms $%
\theta _{\nu }^{A}(x)$ transform like a co-vector we find from the relation
\ Equ. (16) that the vector components $v^{M}$ in the moving frame are
invariant with respect to general coordinate transformations $%
v^{M}=v^{\prime M},F^{M}=F^{\prime M}.$ Therefore the operators $div_{v}$
and $\Delta _{v}$ in the last two terms o.r.s. of Equ.(39) are also
invariant. From the transformation described by Equ. (41) we find with $%
x^{0}=x^{\prime 0}$
\begin{equation}
\frac{\partial }{\partial x^{i}}=\frac{\partial f^{j}}{\partial x^{i}}\frac{%
\partial }{\partial x^{\prime j}}.  \label{Equ.(43b)}
\end{equation}

The first term o.r.s. of Equ.(39) including the three-dimensional $div_{x}$
operator with respect to the 3-geometry of the hypersurface is intrinsically
defined by the hypersurface and therefore invariant under the
hypersurface-preserving transformation (41). This can be simply proven by
the transformation property of the intrinsic covariant differentiation $%
^{(3)}\Delta _{i}F^{j}$ on the hypersurface given by $^{(3)}\Delta
`_{j}F`^{i}=(\partial x`^{i}/\partial x^{k})(\partial x^{l}/\partial
x`^{j})^{(3)}\Delta _{l}F^{k}$. For the divergence operator $div_{x}(\mathbf{%
F}\phi )=$ $^{(3)}\Delta _{j}(F^{j}\phi )$ this yields with $(\partial
x`^{i}/\partial x^{k})(\partial x^{l}/\partial x`^{i})=\delta _{l}^{k}$ the
relation $^{(3)}\Delta `_{j}F`^{j}$ =$^{(3)}\Delta _{j}F^{j}$ or $%
div_{x^{\prime }}(\mathbf{F}^{\prime }\phi )=div_{x}(\mathbf{F}\phi ).$ The
time derivative on the left-hand side transforms under Equ.(41) as

\begin{eqnarray}
\frac{\partial \Phi }{\partial x^{0}} &=&\frac{\partial \Phi }{\partial
x^{\prime 0}}-\beta ^{k}\frac{\partial }{\partial x^{\prime k}}\Phi
\label{Eaq.(42b)} \\
\beta ^{k} &=&\frac{\partial f^{k}}{\partial x^{j}}(\frac{\partial x^{j}}{%
\partial x^{0}})_{f^{k}=const}  \notag
\end{eqnarray}

The transformed relativistic Kramers equation (39) takes therefore
the form :

\begin{eqnarray}
N^{-1}v^{\prime 0}(\frac{\partial \Phi }{\partial x^{\prime 0}}-\beta ^{k}%
\frac{\partial }{\partial x^{\prime k}}\Phi ) &=&-v`^{A}div_{x\prime }(%
\mathbf{e}_{A}^{\prime }(x)\Phi )  \label{Equ. (43a)} \\
&&-div_{v\prime }(\mathbf{F}\Phi )+\frac{D}{2}\Delta _{v\prime }\Phi ,
\notag
\end{eqnarray}%
where
\begin{equation}
div_{x^{\prime }}(\mathbf{e}_{A}^{\prime }(x^{\prime })\Phi )=\frac{1}{\sqrt{%
g^{\prime }}}\frac{\partial }{\partial x^{\prime i}}(\sqrt{g^{\prime }}%
e_{A}^{\prime i}(x^{\prime })\Phi ),  \label{Eq.(45)}
\end{equation}%
and $g^{\prime }=gJ^{2}$ where $J$ is the Jacobian matrix of the
transformation: $J=\det \{\frac{\partial f^{j}}{\partial x^{i}}$\}. A time
rescaling Equ.(42) do not change the left-hand side of Equ.(39) \ because
the entity $N$ is transformed like $N^{\prime }=N(\partial t/\partial
t^{\prime })$and therefore ($N)^{-1}\partial /\partial t=(N^{\prime
})^{-1}\partial /\partial t^{\prime }.$

Note that transformation properties as found in Equ. (45) with the
replacement of the time derivative by the left-side in Equ.(45) is a general
feature of the evolution formalism in the 3+1 spacetime decomposition in the
general relativity theory (see e.g. \cite{Bona}).

\section{5. Diffusion in the expanding universe}

Let us discuss the above given relativistic diffusion equation (39) in
gravitational fields for the example of the expanding universe. The metric
that characterize the expanding spatial homogenous and isotropic universe
can be described by the Robertson-Walker metric which has the form%
\begin{equation}
ds^{2}=(cdt)^{2}-\frac{R(t)^{2}}{1-\varepsilon r^{2}}%
[(dx^{1})^{2}+(dx^{2})^{2}+(dx^{3})^{2}],  \label{Eq.(46)}
\end{equation}%
where $\varepsilon $ may assume the values 0, 1or -1 for a flat, closed or
open universe, respectively, r$^{2}=(x^{1})^{2}+(x^{2})^{2}+(x^{3})^{2}$ and
$R(t)$ is the cosmic scale factor. In the following we restrict ourself to
the spatial flat case with $\varepsilon =0.$ With the metric coefficients $%
g_{00}=-1,g_{0i}=0,g_{ij}=-R(t)^{2}\delta _{ij}$ we find for the Christoffel
coefficients \cite{r17}, \cite{r27}

\begin{equation}
\Gamma _{jk}^{i}=0,\Gamma _{jk}^{0}=R\frac{\partial R}{\partial x^{0}}\delta
_{jk},\Gamma _{0k}^{i}=\frac{\partial R}{\partial x^{0}}R^{-1}\delta _{ik}.
\label{Eq.(47)}
\end{equation}%
The moving frames related with the metric in Equ.(47) are given by
\begin{equation}
e_{j}^{a}=R(x^{0})^{-1}\delta _{j}^{a},e_{0}^{0}=1,\theta
_{a}^{j}=R(x^{0})\delta _{a}^{j},\theta _{0}^{0}=1.  \label{Eq.(48)}
\end{equation}%
Using Equ.(48) and (49) we find for the spin connection coefficients

\begin{equation}
\Omega _{jb}^{a}=0,\Omega _{j0}^{a}=\frac{\partial R}{\partial x^{0}}\delta
_{j}^{a},\Omega _{0b}^{a}=0.  \label{Eq.(49)}
\end{equation}%
The gravitational force $F_{g}^{a}=$ $-\Omega _{\mu B}^{a}(x)e_{C}^{\mu
}(x)u^{B}u^{C}$ is given by

\begin{equation}
F_{g}^{a}=-\frac{\partial R}{\partial x^{0}}R^{-1}u^{a}u^{0}.
\label{Eq.(50)}
\end{equation}

\bigskip Let us discuss the diffusion of massive particles in the expanding
universe. Introducing the hyperbolic coordinate system for the 4-velocity,
defined by $u^{1}=\sh\alpha \sin \vartheta \cos \varphi ,$ $u^{2}=\sh\alpha
\sin \vartheta \sin \varphi ,$ $u^{3}=\sh\alpha \cos \vartheta $ and $u^{0}=%
\ch\alpha $ we find for the gravitational force F$_{g}^{\alpha }(\eta
)=-H(\eta )sh\alpha ,$ $F_{g}^{\vartheta }=F_{g}^{\varphi }=0$ where $H(\eta
)=$ $\frac{\partial R}{\partial \eta }R^{-1}$ is the Hubble constant and $%
\eta =x_{0}=\tau ch$. The random impact of surrounding particles generally
cause two kind of effects: they act as a random driving force leading to a
random motion and they give rise to a frictional force. In the
non-relativistic theory the friction force is given by $f_{F}^{i}=-\nu
mv^{i} $ where $\nu $ is the friction coefficient and $v^{i}$ are the
components of the non-relativistic velocity. The relativistic generalization
of the friction force requires the introduction of a friction tensor $\nu
_{\alpha }^{i}$ similar to the pressure tensor in the relativity theory \cite%
{r6}, \cite{r8}. The friction force is expressed as $F_{F}^{i}=\nu _{\alpha
}^{i}[u^{\alpha }-U^{\alpha }],$ where $U^{\alpha }$ is the 4-velocity of
the heat bath. For an isotropic homogeneous heat bath the friction tensor is
given by

\begin{equation}
\nu _{\alpha }^{i}=\nu (\eta _{\alpha }^{i}+u^{i}u_{\alpha }),
\label{Equ.21}
\end{equation}%
with $\nu $ denoting the scalar friction coefficient measured in the rest
frame of the particles. In the laboratory frame the heat bath is at rest
described by $U^{\alpha }=(1,0,0,0)$. Therefore the friction force is given
by $F_{F}^{i}=-\nu u^{i}u^{0}$ or in hyperbolic coordinates%
\begin{equation}
F_{F}^{a}=-\nu \sh\alpha ,F_{F}^{\vartheta }=F_{F}^{\varphi }=0.
\label{Eq.friction}
\end{equation}

In the expanding universe the coefficients $\nu (\eta )$ and $D(\eta )$
depend on time and it is convenient to use the diffusion equation (39) in
term of the\ parametrization with the observer time $\eta $. In the case of
a spatial homogenous and isotropic solution in Equ.(39) the spatial
derivatives vanish. Substituting the ansatz $\ \phi =\phi _{M}^{J}(\alpha
,\vartheta ,\varphi ,\eta )=g_{J}(\alpha ,\eta )Y_{M}^{J}(\theta ,\varphi )$
with $Y_{M}^{J}(\theta ,\varphi )$ =P$_{M}^{J}(\vartheta )$e$^{iM\varphi }$
as the spherical harmonics and the associated Legendre functions P$%
_{M}^{J}(\vartheta )$ into Equ. (39) the following equation can be derived
for $g_{J}=g_{J}(\alpha ,\eta ):$

\begin{eqnarray}
ch\alpha \frac{\partial }{\partial \eta }g_{J} &=&[\frac{D(\eta )}{2}(\frac{%
\partial ^{2}}{\partial \alpha ^{2}}+2\cth\alpha \frac{\partial }{\partial
\alpha }-\frac{J(J+1)}{\sh^{2}\alpha })+  \notag \\
&&+\chi (\eta )(\sh\alpha )^{-2}\frac{\partial }{\partial \alpha }\sh\alpha
)^{3}]g_{J}.  \label{Eq.(54)}
\end{eqnarray}%
with $\chi (\eta )=\frac{\partial R}{\partial \eta }R^{-1}+\nu (\eta ).$
Here the discrete index $J$ takes the values $J=0,1,2,\ldots $ and $%
M=-J,-J+1,\ldots 0,1,\ldots J$. For the fundamental solution $J=0$ we find
the diffusion equation

\begin{eqnarray}
ch\alpha \frac{\partial }{\partial \eta }g_{0} &=&[\frac{D(\eta )}{2}(\frac{%
\partial ^{2}}{\partial \alpha ^{2}}+2\cth\alpha \frac{\partial }{\partial
\alpha })+  \notag \\
&&+\chi (\eta )(\sh\alpha )^{-2}\frac{\partial }{\partial \alpha }\sh%
^{3}\alpha ]g_{0}.  \label{Eq.(55)}
\end{eqnarray}%
\bigskip Let us discuss the solution of this equation within a certain range
of validity substituting the special ansatz
\begin{equation}
\phi (\alpha ,\eta )=C\exp \{\beta (\eta )-\gamma (\eta )\ch\alpha \}
\label{Eq.(52)}
\end{equation}%
into Equ.(55) which yields the relation
\begin{eqnarray}
\frac{\partial \beta }{\partial \eta }-\frac{\partial \gamma }{\partial \eta
}ch\alpha &=&\chi (\eta )[3-\gamma sh\alpha th\alpha ]  \label{Eq.(53)} \\
&&+\frac{D(\eta )}{2}(-3\gamma +\gamma ^{2}sh\alpha th\alpha ).  \notag
\end{eqnarray}%
In general, there do not exist functions $\beta (\eta )$ and $\gamma (\eta )$
which solve this equation, but if we restrict ourself to the
ultrarelativistic case $\alpha \gg 1$ we find as equations for the
coefficients $\beta (\eta )$ and $\gamma (\eta )$:
\begin{eqnarray}
\frac{\partial \gamma }{\partial \eta } &=&\chi (\eta )\gamma -\frac{D(\eta )%
}{2}\gamma ^{2},  \label{Eq. (54)} \\
\frac{\partial \beta }{\partial \eta } &=&3\chi (\eta )-\frac{3D(\eta )}{2}%
\gamma .  \notag
\end{eqnarray}

\bigskip\ The constant $\beta $ is included into the normalization of $\phi
(\alpha ,\eta )$. An analytical solution of Equ.(58) for $\gamma (\eta )$
can be found by a transformation of the variable $\gamma (\eta )=(Y(\eta
))^{-1}$ which yield the solution%
\begin{equation}
Y=Y_{0}\frac{R_{0}}{R(\eta )}e^{-h(\eta )}+\frac{1}{2}\frac{1}{R(\eta )}%
e^{-h(\eta )}\int\limits_{\eta _{0}}^{\eta }e^{h(t^{\prime })}R(t^{\prime
})D(t^{\prime })dt^{\prime },  \label{Eq. (55)}
\end{equation}%
with $h(t)=\int\limits_{t_{0}}^{t}\nu (t^{\prime })dt^{\prime }.$

By comparing the J\"{u}ttner distribution with the solution ansatz (56) we
can introduce a time-depending temperature of the expanding universe%
\begin{equation}
\gamma (\eta )=mc^{2}(kT(\eta ))^{-1}.  \label{Eq. (56)}
\end{equation}%
Since the diffusion and drift constants $D$ and $\nu $ are determined by the
scattering processes of the particles in the system described by different
physical parameters; in particular by the temperature, the solution (59) has
for temperature depending friction and diffusion coefficients the meaning of
an integral equation. However, we have to taken into account that the
diffusion and friction coefficients are related each others by the
fluctuation-dissipation theorem. This relation is well known in the
non-relativistic case where the viscous friction coefficient $\nu $ of a
Brownian particle must be related to the diffusion constant $D$ of the
particles by the Einstein relation

\begin{equation}
D=\frac{2\nu kT}{mc^{2}}.  \label{Eq. (57)}
\end{equation}

The nature of the random force is independent on the presence of the
gravitational field. Therefore in the relativistic case the stationary
solution of the Equ. (55) for $\frac{\partial R}{\partial \eta }%
R^{-1}\rightarrow 0$ must coincide with the J\"{u}ttner distribution. From
the recently derived special relativistic diffusion equation \cite{r27} it
was shown that the J\"{u}ttner distribution arise as the stationary solution
of the special relativistic diffusion equation (9) if the Einstein relation
is not only valid in the non-relativistic case, but also in the relativistic
regime. Substituting the relation (61) with (60) into Equ.(58)\ one can see,
that the effect of diffusion is canceled by the viscosity. Then\ from
Equ.(58) we obtain%
\begin{equation}
\gamma (\eta )=\gamma _{0}\frac{R(\eta )}{R_{0}}.  \label{Eq. (58)}
\end{equation}%
This relation is identical with the result in the kinetic theory for the
expanding universe \cite{r27}, \cite{r29}. In the radiation dominated period
in a flat cosmos we have $R(\eta )\sim \sqrt{\eta }$ and in the matter
dominated period $R(\eta )\sim \eta ^{2/3}$ and therefore we find for the
ultra-relativistic case $T\sim \eta ^{-1/2}$ or $T\sim \eta ^{-2/3},$
respectively for the corresponding periods.

Note that the solution (56) do not satisfy physically determined initial
conditions; this solution describes the asymptotic quasi-static regime and
is valid only after a certain time when the system is already in the
equilibrium. As shown above, in this case the diffusion is compensated by
the drift process, and the result that follows is consistent with the
equilibrium state in kinetic theory. On the other hand in the opposite
transient case up to a certain time $\eta $ after the initial time $\eta
_{0} $ one can neglect in Equ.(55) the last term proportional to $\chi (\eta
).$ Then in the ultra-relativistic case the fundamental solution $J=0$ is
determined by the equation%
\begin{equation}
\frac{\partial }{\partial \varsigma }\Phi =\frac{1}{2}(\xi \frac{\partial
^{2}}{\partial \xi ^{2}}+3\frac{\partial }{\partial \xi })\Phi ,
\label{Eq. (59)}
\end{equation}%
where the new variables $\varsigma =\int\limits_{\eta _{0}}^{\eta }D(t)dt$
and $\xi =\exp (\alpha )$ are introduced. Using the Laplace transformation $%
\Phi (\varsigma ,\xi )=\int\limits_{0}^{\infty }\widetilde{\Phi }(\lambda
,\xi )\exp (-\lambda \varsigma )d\varsigma $ we find the solution%
\begin{equation}
\widetilde{\Phi }(\lambda ,\xi )=\xi ^{-1}J_{1}(2\sqrt{\xi \lambda }),
\label{Eq. (60)}
\end{equation}%
where $J_{1}$ is the first order Bessel function. The eigenfunctions $%
\widetilde{\Phi }(\lambda ,\xi )$ satisfy the relations of orthogonality.
Therefore the transition probability is determined by

\begin{equation}
\Phi (\xi ,\varsigma \mid \xi _{0,}0)=\int\limits_{0}^{\infty }\widetilde{%
\Phi }(\lambda ,\xi \ )\widetilde{\Phi }^{\ast }(\lambda ,\xi _{0})\exp
(-\lambda \varsigma )d\lambda .  \label{Eq. (61)}
\end{equation}%
Substituting Equ.(64) into \bigskip Equ. (65) we find%
\begin{equation}
\Phi (\xi ,\varsigma \mid \xi _{0,}0)=C\xi ^{-1}\varsigma ^{-1}I_{1}(\frac{2%
\sqrt{\xi \xi _{0}}}{\varsigma })\exp (-\frac{\xi +\xi _{0}}{\varsigma }),
\label{Eq. (62)}
\end{equation}%
where C is the normalization constant and $I_{1}(x)=-iJ_{1}(ix).$ In the
general case of arbitrary time the solution of Equ. (54) can be obtained by
numerical methods. But for the study of this problem under the conditions of
the earliest epoch of the universe we need a realistic microscopic model for
the viscosity in the non-equilibrium epoch in a plasma of relativistic
particles, including quarks, leptons, gauge and Higgs bosons. A detailed
discussion of this issue is beyond the scope of the present paper.
Corresponding cosmological estimations the universe may not have been in
thermal equilibrium during its earliest epoch in a time range earlier than
about $10^{-38}s$ after the big bang or temperatures greater than $%
10^{16}GeV $ \cite{r29}. Standard phenomenological inflationary cosmological
models relate this epoch with symmetry breaking phase transitions.

Let us finally briefly discuss the non-relativistic limit $\alpha \ll 1.$%
From Equ.(57) we find under this condition:
\begin{eqnarray}
\frac{\partial \gamma }{\partial \eta } &=&2\chi (\eta )\gamma -\ D(\eta
)\gamma ^{2},  \label{Eq. (63)} \\
\frac{\partial \beta }{\partial \eta } &=&-\frac{\partial \gamma }{\partial
\eta }+3\chi (\eta )-\frac{3D(\eta )}{2}\gamma .  \notag
\end{eqnarray}

If we use the Einstein relation (61) the diffusion term is again canceled by
the friction and from Equ.(67) the solution $\gamma (\eta )=\gamma _{0\text{
}}R^{2}(\eta )/R_{0}^{2}$ follows. The same solution is obtained in the
kinetic theory for a non-relativistic gas. In the non-relativistic case the
temperature of the equilibrium distribution therefore depends on the cosmic
scale factor $R(\eta )$ like $T\sim R^{-2}(\eta ).$ Thus, both in the
relativistic and in the non-relativistic case the validity of the general
fluctuation-dissipation theorem with the Einstein relation (61) ensures that
the result of the kinetic theory derived from the vanishing of the collision
integral in the Boltzmann equation is consistent with the here derived
probabilistic general relativistic diffusion theory in the quasi-steady
state regime or in thermodynamic equilibrium.

\section{6. Conclusions}

In conclusion, a theory of Markovian diffusion processes within the
framework of the general theory of relativity is formulated. In the
derivation of the basic relativistic diffusion equation the mathematical
calculus of stochastic differential equations on Riemannian manifolds is
used, which here is modified for the description of diffusion in the phase
space of Pseudo-Riemanian manifolds with an indefinite metric by using
orthonormal frame vectors both in the position and in the velocity space. A
generalized Langevin equation in the fiber space of position, velocity and
orthonormal velocity frames is defined and the generalized Kramers equation
within the framework of general relativity is derived both in the
parametrization of the phase-space proper time and the observer time. The
transformation of the obtained diffusion equation under
hypersurface-preserving coordinate transformations is studied and diffusion
in the expanding universe is discussed. It is shown that the validity of the
fluctuation-dissipation theorem in the relativistic case ensures that in the
quasi-steady state regime the result of the derived diffusion equation is
consistent with the kinetic theory in thermodynamic equilibrium. Besides a
transient analytical solution valid for small times has been derived.



\begin{thebibliography}{99}
\bibitem{r1} J. Lopuszanski, Acta Phys. Pol. \textbf{12,} 87 (1953).

\bibitem{r3} R. M. Dudley, Ark. Mat. Astron, Fys. \textbf{6}, 241 (1965).

\bibitem{r4} R. Hakim, J. Math. Phys. \textbf{6}, 1482 (1965)

\bibitem{r6} F. Debbasch and J. P. Rivet, J. Stat. Phys. \textbf{90}, 1179
(1998).

\bibitem{r8} J. Dunkel and P. H\"{a}nggi, Phys. Rev. E \textbf{72}, 036106
(2005).

\bibitem{r9} O. Oron, L. P. Horwith, Found. Phys. \textbf{35}, 1181 (2005.)

\bibitem{r12} G. Chacon-Ascosta and G. M. Kremer, Phys. Rev. E \textbf{76},
021201 (2007)

\bibitem{r10} Z. Haba, Phys. Rev. E \textbf{79}, 021128 (2009).

\bibitem{r11} J. Dunkel and P. H\"{a}nggi, Physics Reports, \textbf{471}, 1
(2009).

\bibitem{a1} F. Debbasch, J. Math. Phys. \textbf{45}, 2744 (2004)

\bibitem{a2} J. Franchi, Y. Le Jan; Comm. Pure Appl. Math. \textbf{60}, 187
(2006)

\bibitem{a3} C. Chevalier,F. Debbasch, J. Math. Phys. \textbf{48}, 023304
(2007)

\bibitem{r14} F. J\"{u}ttner, Annalen der Physik, \textbf{34}, 856 (1911).

\bibitem{r15} S. R. de Groot, W. A. Leeuwen and Ch. G. Weert, "Relativistic
kinetic theory", North Holland 1980.

\bibitem{r16} C. Cercignani, G. M. Kremer, "Relativistic Boltzmann equation:
Theory and applications", Birkh\"{a}user Verlag, Basel 2002.

\bibitem{k1} W. Israel, J.W. Stewart, Ann. Phys. 118, 341 (1979)

\bibitem{k2} J. M. Stewart, "Non-equilibrium relativistic kinetic theory".
Lecture Notes in Physics, vol.\textbf{10}, Springer, Berlin (1971)

\bibitem{k3} J. Ehlers, General relativity and kinetic theory". In B. K.
Sachs (Ed.) "General Relativity and Cosmology", Proc.of the Int. School of
Physics" Enrico Fermi", Academic Press (1971)

\bibitem{r17} J. Bernstein, "Kinetic theory in the expanding universe",
Cambrudge University Press 1988

\bibitem{d2} R. Martens, Class. Quantum Grav. \textbf{12}, 1455 (1995)

\bibitem{d3} M. K. Mak and T. Harko, Int. J. Mod. Phys. D \textbf{12}, 925
(2003)

\bibitem{k4} M.A. Schweizer, Astron. Astrophys. \textbf{151}, 79 (1985)

\bibitem{k5} I.S. Liu, I M\"{u}ller, T. Ruggeri, Ann. Phys. (NY) \textbf{169}%
, 191 (1986)

\bibitem{k7} W. Hu, D. Scott and J. Silk, Phys. Rev. D \textbf{49}, 648
(1994)

\bibitem{k6} Z. Banach, Physica A \textbf{275}, 405 (2000)

\bibitem{a4} C-M Ma and E.Bertschinger, Astophys. J. \textbf{612}, 28 (2004)

\bibitem{a5} D. Meritt, Astrophys. J. \textbf{568}, 998 (2002)

\bibitem{a6} V. Berezinsky, A. Z. Gazizov, Astrophysical Journal, \textbf{643%
}, 8 (2006)

\bibitem{r24} N. Ikeda and S. Watanabe, "Stochastic differential equations
and diffusion processes", North Holland 1989.

\bibitem{r25} E. P. Hsu, "Stochastic analysis on manifolds", Graduate
Studies in Mathematics, American Mathematical Society, Providence, Rhode
Island.

\bibitem{r26} J. Herrmann, Phys. Rev. E \textbf{80}, 051110 (2009), arXiv:
0903.0751v1

\bibitem{r27} C. W. Misner, K. S. Thorne, J. A. Wheeler, "Gravitation", W.
H. Freeman and Company, 1973

\bibitem{r28} Bo Yuan Hou, "Differential geometry of physicists", World
Scientific Publishing, 1997

\bibitem{Bona} C. Bona et al.: "The Evolution Formalism", Lecture Notes in
Physics \textbf{783}, 25-48 (2009)

\bibitem{r29} E. W. Kolb and M. S. Turner, "The early universe",
Addison-Wesley Publishing Copany, 1990
\end{thebibliography}
\end{document}